# Implications for Utilizing YouTube based Community Interactions for Destination Marketing

## Investigation of a Typology Approach

Arunasalam Sambhanthan[1], Samantha Thelijjagoda[2]
Department of Information Management
Sri Lanka Institute of Information Technology
Malabe, Sri Lanka
arunsambhanthan@gmail.com[1], samantha.t@sliit.lk[2]

Joseph Tan[3]
DeGroote School of Business
McMaster University
Ontario, Canada
tanjosep@mcmaster.ca[3]

*Abstract*— In recent time, YouTube has evolved into a powerful medium for social interaction. Utilizing YouTube for enhancing marketing endeavors is a strategy practiced by marketing professionals across several industries. This paper rationalizes on the different ways and means of leveraging YouTube-based platforms for effective destination marketing by the hospitality industry (hotels). More specifically, the typology of virtual communities is adapted to evaluate the YouTube platform for effective destination marketing. Comments made by YouTube users have been subjected to a content analysis and the results are reported here under the five broad clusters of virtual communities. Implications for utilizing YouTube-based community interactions for destination marketing are also highlighted as part of the outcome of this research.

*IndexTerms*—YouTube, Destination Marketing, Virtual Communities, Interaction Design, Tourism

## 1. Introduction

A plethora of literature and published materials exists on marketing destinations via the use of social media [1]. Even so, there still is a paucity of research on YouTube-based Virtual Communities for destination marketing [2]. Evidently, the use of YouTube as a medium to market tourist destinations remains an open question for exploration in the context of rapidly emerging marketing models using web 2.0. The investigation presented here uses a typology-based approach to assess the interaction effectiveness of YouTube-based virtual communities. As well, the study focuses on developing a set of implications for utilizing YouTube-based community interactions as a means for destination marketing.

## 2. Virtual Communities

Romm *et al.* defined virtual communities as "the groups of people who communicate with each other via electronic media" [3]. While this definition shows the potential for a wide-ranging scope of virtual communities to be formed; the scope of the current research effort has been restricted primarily to virtual communities that are based on YouTube platform with its context focused on destination marketing. In this context, the term virtual community (VC) is redefined as "the groups of people" who communicate tourism-related destination information and experiences with each other via YouTube.

### 2.1. Typology of Virtual Communities

The conceptual framework for this research is based on the typology of virtual community [4]. The typology of virtual communities outlines five major themes under which these communities could be studied. The five main themes are derived from the 5Ps of virtual communities. Messinger, *et. al.* discussed the typology for Virtual Communities, and the historical developments of research in Social Virtual Worlds [5]. The authors clearly outlined the development of the gaming industry as well as the social networking industry while explaining the five (5) Ps typology of a VC as propounded by prior research: Purpose, Place, Platform, Population, and Profit Model. Additionally, the Messinger, *et. al.* paper uses the typology to interpret the historical progression of innovations in the electronic gaming and social networking industries. The authors argued that the typology could assist in identifying the future applications for business, e-commerce and education, as well as potential new technological features and research topics in social, business and computing sciences. The above claim could be accepted in specific cases only. Especially, the typology has been originally proposed for classifying the virtual communities. The elements of the typology indicate aspects such as population which is very much contextual to virtual communities. However, the other aspects such as purpose, place, platform and profit model could be applicable to almost all other domains. Therefore, the above claim could be accepted under the grounds that the claim is made based on a generalization of aspects which could be applied to other domains than virtual communities. However, the applicable domain needs to be scientifically tested in order to derive any meaningful conclusions on the above. This research, having been applied to the use of YouTube based virtual communities for destination marketing, aims to test the above claim for its e-commerce related applications.

The Messinger, *et. al.* typology, which is adapted in guiding the thinking of this research, is outlined below:

1. **Purpose:** Focuses on the specific type of information or content being communicated among the VC. Content delivery specific to a case could be strategic, managerial or tactical in nature. In this research we are using this term to denote the focus, effectiveness and accuracy of the content being presented through YouTube based discussions. The focus of user generated content in the YouTube, the effectiveness of information presented by the users during the discussions and the accuracy of that content are evaluated under the theme of purpose in this research.

2. **Place:** Focuses on whether the location is completely virtual, or only partially virtual. The nature of place would depend on whether the participants are co-located or geographically dispersed. The geographic locations of the YouTube users and the trusting behavior among the users are focused under this section. The term place has been designated through the geographic location as well as the trusting behavior, which is a form of psychological nearness among users. Therefore the term place used in the form of physical and psychological locations of the users of YouTube platform.

3. **Platform:** Focuses on the type of communication involved during the interaction, whether it is synchronous and/or asynchronous communication, or both. The term platform has been designated to evaluate the design of communication medium utilized for YouTube based interaction among the users. This section evaluated the effectiveness of synchronized communication medium among the users.

4. **Population:** Focuses on the size of the participating interacting group. As social ties have less connectivity with commercialization, distinguishing target user market characteristics would be more critical here, although Porter (2004) also considered the types of social ties. In this section, we are evaluating the size of the population and the social ties among the user groups. The section also focuses on the psychological attachment of users which could be utilized as a means for destination marketing.

5. **Profit Model:** Focuses on revenue or non-revenue generating environments. Here, a taxonomy of the profit model can range from: (1) a fixed fee model like a single purchase price or registration fee; (2) a variable fee model such as a fee per use; (3) a subscription-based model (and on what basis subscriptions are being made); (4) an advertising-based model; (5) a pay-as-you-go model (e.g., payable extras for virtual assets including clothing, land, and software); and (6) an ancillary sale model such as the sale of products like real stuffed animals and accessories, which are accompanied by passwords for accounts in Virtual Worlds where virtual versions of the products enable combined real and virtual play by interacting participants. This section has been designated for evaluating the proposed profit model which could be adapted by destination marketers for exchanging ideas, products and experiences related to tourism.

3. METHODOLOGY

A content analysis of YouTube videos was conducted for this research. A search on Sri Lankan Tourism has been conducted in the YouTube search window. Five videos were randomly selected for further evaluation. The user group include novice users, frequent users, proficient and knowledgeable users as well as expert users. However, the users have not been categorized into any of the above classification as the aim of the research was to evaluate the applicability of YouTube based community interactions as a way forward for destination marketing, which does not really needs to have the category of users. Arguably, the tourism experience is constant for all users regardless of the type they fall in. We studied 137 user comments from the YouTube platform were subjected to content analysis using the five Ps typology propounded by Porter (2004). Five main themes were utilized to categorize the results and build the knowledge base for this research. The 5Ps has been propounded by Porter as a way forward for classifying virtual communities into a typology. Messinger et. al utilized these typologies to evaluate the development of social networking and gaming industries. We, in this research utilize the same 5Ps to evaluate the applicability of YouTube platform for destination marketing. The application at this point is generalized into the suitability of 5Ps typology for evaluating the e-commerce domain.

4. RESULTS

**4.1. Purpose:** Focuses on the specific type of information or content being communicated among the VC. Content delivery specific to a case could be strategic, managerial or tactical in nature. In this research we are using this term to denote the focus, effectiveness and accuracy of the content being presented through YouTube based discussions. The focus of user generated content in the YouTube, the effectiveness of information presented by the users during the discussions and the accuracy of that content are evaluated under the theme of purpose in this research.

### 4.1.1. Content Focus

In analyzing the focus of the content, participating users have been found to be discussing about the Sri Lanka weather before vacationing in the country.

"I might go to Sri Lanka next month because I want to go somewhere nice to film, I wanted to go to Spain because I know it will be lovely, sunny and bright for filming but I will be going on my own and I need wheels to get about but I don't have a license, but I'm guessing you don't need a license to hire a moped in Sri Lanka. My only fear is it will rain a lot in Sri Lanka in June"(**User 14**)

The above comment shows that the prospective tourists had expressed interest in visiting to Sri Lanka, but were uncertain about the likely change in climate in the country over a specific time period. Also, the commentary shows that these tourists did have alternatives, which could be selected for satisfying their specific tourism needs. Moreover, the comment reveals that the person was interested in filming the destination and was really looking for a place where the natural beauty is available. An interesting observation made by another participant interacting in the above discussion shows the participative nature of the YouTube users and their willingness to support future tourists to their home country.

"Great Video..! Rain is usually trouble during April, May, October and November. Other than that Sri Lanka is Sunny." **(User - 5)**

In developing countries such as Sri Lanka, our analysis simply shows that community-based interactions could be effectively capitalized for destination marketing. Importantly, the enhancement of community-based interactions with a business focus could effectively form an encouraging strategy for tourism promotion through the content of interaction.

### 4.1.2. Content Effectiveness / Relevance

In analyzing the effectiveness of the video content, users have been found to be discussing about several points in the YouTube platform which are not relevant for tourism destination marketing. In particular, some of the discussions observed were unrelated to tourism but showed the interpersonal biases of the users, for example, when impromptu remarks and/or opinions were exchanged on matters such as politics.

"Nice country but stupid greedy politicians eating everything in their sight. Waste" **(User 23)**

Although the aforementioned content has been observed in one of the videos, there is an interesting aspect in that users were making comparative suggestions about the tourism destinations.

"I agree that Hawaii is way too overrated!! The reason why people go to Hawaii is its closer, cheaper to travel to, and most people don't know what SriLanka is like" **(User – 29)**

"That's just plain snobbery. Give credit when it is due. This is not a competition of who/what/where is the best. It is a short clip that showcases the beauty of Sri Lanka in 2 minutes. Learn to appreciate life instead of having the need to make comparisons all the time. You'll be happier too." **(User – 30)**

The above content shows that the users were trying to make comparative assessment of different destinations before making a travel plan. This is a critical point whether the purchase decision of tourism products were made by the users. The destination marketing aspirations of hotels should focus on this point, as this is a pre-consumption decision point.

Indeed, the nature of comparative thinking of the consumers could pose challenges to destination marketing through the YouTube platform. Arguably, the users will tend to compare the tourism products and would tend to adapt the high quality product which has more advantages than what is offered by destination marketers. In addition, the power of user generated content via the YouTube platform is another bottleneck for effective tourism promotion. In fact, an unsatisfied user will definitely spread the word through this kind of platform, which will create a virtual thread for destination marketer through negative electronic word of mouth (e-WOM) .Although, these discussions could be positively utilized for any destination marketing efforts, the challenge of tourists shifting to other products is considerably high with regard to this scenario. Therefore the YouTube designers should be afraid of the effectiveness of content which is presented through the YouTube platform due to the fact that the naturally users are empowered with user generated content for discussing their ideas and experiences about products in the cyberspace. Hence, effective design of YouTube content is crucial for the success of any sorts of efforts towards destination marketing.

### 4.1.3. Content Accuracy

The next concern to be addressed is the accuracy of content presented through the YouTube platform. A user comment made on Bikinis is an example for how the accuracy of information could affect the destination marketing of hotels located in the developing nations. Especially, inaccurate information presented by the fellow users could possibly mislead the users in tourism related consumption decisions.

"Bikinis were never banned in Sri Lanka! Are you studying law? Then you would know that; what you are playing here is a hoax. Only mini-skirts were to be banned in Sri Lanka although with the growth of the tourist population this rule was not imposed! Please check your facts before you comment on a video on your country." **(User - 6)**

This commentary clearly shows how the users are interacting and debating about tourism-related information. In fact, it also implies that the misleading information in the cyberspace could influence either positively or negatively on users decisions. This needs to be affectively tackled in order to make a positive impact on destination marketing. The video owners should have a regular review on the user generated content in the YouTube to ensure that the users are not given with misleading information by fellow users. In fact, there should be some mechanisms to moderate the information presented in the comments section to ensure relevant and accurate information is presented in the discussion section.

**4.2. Place:** Focuses on whether the location is completely virtual, or only partially virtual. The nature of place would depend on whether the participants are co-located or geographically dispersed. The geographic locations of the YouTube users and the trusting behavior among the users are focused under this section. The term place has been designated through the geographic location as well as the trusting behavior, which is a form of psychological nearness among users. Therefore the term place used in the form of physical and psychological locations of the users of YouTube platform.

### 4.2.1. Geographic Diversity

Overall, the articulations show that the users from different countries are taking part in the discussions. As for the type of discussions taking place in the YouTube platform, our analysis shows that several topics would typically be discussed in the portal. Essentially, the products types used to video the destination were debated among the users. Also the users are having a healthy discussion on the beauties of the destinations as well as the history.

"This is the most amazing video of Taj Mahal. I just would like to ask you which video camera have you used? It's as clear as blue ray. Please tell me which video camera have you used? Love from India" **(User - 33)**

"We used a canon 60d. Cheers" **(User - 34)**

The above discussion shows that the products types used to record the destination was discussed by the users. It signals that the users have been influenced tremendously by the quality of destination video presented through YouTube. Hence, the quality of video could be featured as an aspect which increases the destination marketing possibilities through the YouTube platform.

### 4.2.2. Trusting Behavior

Critically, the trust between users would have a tremendous impact on the destination marketing. Specifically, the trust between users will have a direct influence over the dependability they will have in the comments that were made by others and which, in turn, would have an impact on how much they relied on the information provided by the other user.

"Doubtful. It was dark and full of people. The center is fenced off so no one could even get close to the marble cenotaphs. Videos/photos are forbidden, but the inside is not nearly as pretty as the outside anyway." **(User - 52)**

This comment shows that the user is trying to recall a part of his memories about the same location to validate the dependability and reliability of the content that was uploaded by the other user. It further shows that users have a tendency of validating the virtual experience with the actual travel experience; in fact, such action will direct them to either rely upon or reject to some degree the virtual experience provided through YouTube. As noted, user trusting behavior is critical for effective destination marketing and most, if not all, users tend to make judgments about the uploaded content by comparing it against their original experience.

**4.3. Platform:** Focuses on the type of communication involved during the interaction, whether it is synchronous and/or asynchronous communication, or both. The term platform has been designated to evaluate the design of communication medium utilized for YouTube based interaction among the users. This section evaluated the effectiveness of synchronized communication medium among the users.

### 4.3.1. Design of Communication Medium

The design of communication medium plays a critical role if one is to effectively utilize the YouTube-based interactions for destination marketing. Typically, the YouTube platform encourages a synchronous communication through enabling two-way communications among users.

"Excellent photos"**(User – 35)**

"Thank You!" **(User - 36)**

"Well-put together video! One day I'll have someone to travel the world with!" **(User - 37)**

"Good luck :)" **(User - 38)**

"This make me want to watch Aladdin" **(User – 39)**

The above series of exchange, which could be effectively utilized for encouraging tourism-related discussion, shows a pattern of positive dialogue among the participants. Particularly, the travel experience of specific destinations could be discussed by the participants through whom the destination marketer could have a grasp on what are the positive/negative aspects which could encourage/discourage destination consumption pattern of users.

In contrast, it is also observed that users have a tendency to enjoy the background music used in the videos.

"The music is so wrong, could be more fun" **(User - 73)**

"Fantastic photos! Could you please tell me which music is being played in the video? Thanks" **(User - 86)**

"Almost complete. Yet you forgot to include the most primitive countries yet a cursor to the civilization in the East, such as Burma, Bagan temples. Nice background music" **(User - 125)**

"Love the music so much" **(User - 90)**

Here, the addition of relevant simulating music to the video clip appeared to have encouraged interaction among the users with regards to the discussion of music. In particular, the type of music selected for the video should be chosen with care in order to enhance positive interactions among users.

**4.4. Population:** Focuses on the size of the participating interacting group. As social ties have less connectivity with commercialization, distinguishing target user market characteristics would be more critical here, although Porter (2004) also considered the types of social ties. In this section, we are evaluating the size of the population and the social ties among the user groups. The section also focuses

### 4.4.1. Size of the Population

YouTube-based community interactions on tourism videos are "medium" in size. The word medium means that the size is not too larger as well as not shorter. Although, there is no quantitative measure utilized here, it implies that the size is of a middle one. There are people coming from different countries, but the size is restricted only to the people who have expressed an interest in tourism. There are people with other interests as well but the size is relatively low. Especially, the motive of the group of people who are interacting is restricted to the discussion and sharing of tourism information and experience. Therefore, more targeted messages on YouTube-based communities could entertain further discussion around the topic and eventually could lead to increased destination consumption.

### 4.4.2. Social Ties

An interesting social tie could be observed among YouTube users. For example, the patriotic feeling of users was an observable phenomenon in the content analyzed as part of this research. Particularly, the users are very patriotic when it comes to their home countries and shows a tendency of promoting destinations located in their home countries.

"Nice video!! Youguys are amazing and funny!! Greetings from a Sri Lankan" **(User - 19)**

"I loved Sri Lanka. I deeply want to return, but this time on a vacation. The people are fantastic. Hope to see you soon Sri Lanka" **(User - 24)**

"I'll see you soon mother Sri Lanka" **(User - 27)**

"Go Sri Lanka" **(User - 28)**

"This is so awesome! Thank you both for the visit inside and out :) very nice place! Hope to visit Agra one day :) and congrats for your engagement" **(User - 48)**

"All of you are wrong the most beautiful place is Nepal. You can come and visit" **(User - 71)**

"It's more fun in the Philippines!" **(User - 74)**

"The most beautiful place in the earth is Pakistan. That's why America wants Pakistan." **(User - 75)**

"You must see them! Iceland literally takes your breath away no matter where you are in the country. Everything is just beautiful! No slums or areas to avoid. Even Reykjavik is gorgeous. In New Zealand - a city is a city like anywhere else. The most beautiful part of New Zealand that we saw was Queenstown, which really reminded me of Iceland**." (User - 107)**

"I have always wanted to see Iceland and New Zealand! The pictures I see of both of them are just gorgeous!" **(User - 108)**

Together, the above content clearly shows that the users have an inherent tendency of promoting tourism destinations belonging to their own countries. This could be effectively utilized for tourism destination marketing through YouTube platform. Simply stated, it is notable that users are positive about developing a dialog with other fellow users with regard to the actual tourism experience they have gained from the destinations they have visited. This is another point for consideration when designing a YouTube based virtual community-based strategy for tourism destination marketing.

**4.5. Profit Model:** Focuses on revenue or non-revenue generating environments. Here, a taxonomy of the profit model can range from: (1) a fixed fee model like a single purchase price or registration fee; (2) a variable fee model such as a fee per use; (3) a subscription-based model (and on what basis subscriptions are being made); (4) an advertising-based model; (5) a pay-as-you-go model (e.g., payable extras for virtual assets including clothing, land, and software); and (6) an ancillary sale model such as the sale of products like real stuffed animals and accessories, which are accompanied by passwords for accounts in Virtual Worlds where virtual versions of the products enable combined real and virtual play by interacting participants. This section has been designated for evaluating the proposed profit model which could be adapted by destination marketers for exchanging ideas, products and experiences related to tourism.

In the case of YouTube, the profit model is not through subscriptions or any other related revenue generating efforts. The users are left to interact among themselves and to make business through the platform through any related business activities. However, the users are observed to be promoting the products through the platform. Typically, the users are promoting tourism destinations located in their own countries. Owing to this, the YouTube-based platform could be featured as a place for exchanging ideas and business commodities through community-based interactions among users. Accordingly, tourism businesses could take advantage through the platform by encouraging effective business oriented discussions to be exchanged among the users.

## 5. IMPLICATIONS

A number of implications were formulated as the outcome of this research. The users have specific tendencies and behavioral patterns when they get to interact as a community. Tracing out these behavioral patterns and developing strategies to utilize those behavioral patterns for increased promotional activities would most definitely boost the promotional endeavors of the businesses through YouTube-based community platforms. The following implications have been observed and further developed as part of the outcome of this research.

1. The weather information of the host country needs to be presented through the YouTube video. Supporting information related to the weather is critical and how tourists could manage it needs also to be presented.
2. Tourists are interested in filming the destination and look for ways to do this. The natural beauty could be replicated through the tourism video so as to encourage tourists to film the destinations.
3. Users have an inherent tendency to promote their homelands to others in the VC. Especially, the participative nature of YouTube-based interactions could be utilized to promote tourist destinations.
4. Users tend to want to make comparative assessment of tourism destinations through YouTube platform. Therefore, it is beneficial to include some comparative benefits in the videos uploaded in the YouTube, which will eventually make users to think in a comparative manner. This also poses a challenge for tourism marketing, as the users would be able to shift to the other products on finding better tourism destinations.
5. Proper moderation and mediation of content posted by the users would ensure the accuracy of content presented under the videos.
6. Users have been influenced tremendously by the quality of destination video presented through YouTube. Hence, the quality of video could be featured as an aspect to increase the destination marketing possibilities through the YouTube platform.
7. The users have a tendency of validating the virtual experience with the actual travel experience, which would influence them to either rely or reject the virtual experience provided through the YouTube. This shows that the trusting behavior of users is critical for effective destination marketing and the users tend to make judgments about the uploaded content by comparing it against their original experience.
8. The chains of discussions show a pattern of positive dialogue between participants. This could be effectively utilized for encouraging tourism-related discussion among the participants. Particularly, the travel experience of specific destinations could be discussed by the participants through whom the destination marketer could have a grasp on what are the positive / negative aspects which encourage / discourage destination consumption pattern of users.
9. It is also observed that users have a tendency to entertain the background music used in the videos. This tendency implies that the addition of relevant simulating music to the video clip will encourage interaction among the users with regard to the discussion of music.
10. The motive of the group of people who are interacting is restricted to the discussion and sharing of tourism information and experience. Therefore, more targeted messages on YouTube based communities could entertain further discussion around the topic and eventually could lead to increased destination consumption.